\documentclass{PoS}

\title{Measurement of Longitudinal Single-Spin Asymmetry for W$^\pm$ Production in Polarized Proton+Proton Collisions at STAR}

\ShortTitle{Measurement of $A_L$ for W$^\pm$ boson in Polarized $p+p$ Collisions at STAR}

\author{\speaker{Qinghua Xu}%
       \thanks{For the STAR collaboration.}\\
      School of Physics, Key Laboratory of Particle Physics and Particle Irradiation (MoE), Shandong University, Jinan, Shandong 250100\\
      E-mail: \email{xuqh@sdu.edu.cn}}

\abstract{
The sea quark contribution to the nucleon spin is an important piece for a complete understanding of the nucleon spin structure. The production of W bosons in longitudinally polarized $p + p$ collisions at RHIC provides an unique probe for the sea quark polarization, through the parity-violating single-spin asymmetry, $A_L$. At the STAR experiment, W bosons through the leptonic decay channel $W \to  e \nu $ can be effectively determined with the Electromagnetic Calorimeters and Time Projection Chamber at mid-rapidity. The previous STAR measurements of $A_L$ for W boson production from datasets taken in 2011 and 2012, have provided significant constraints on the helicity distribution functions of $\bar u$ and $\bar d$  quarks. In 2013 the STAR experiment collected $p+p$ data with an integrated luminosity of  about 300 pb$^{-1}$ at $\sqrt s$ = 510 GeV with an average beam polarization of about  $56\%$, which is more than three times larger than the total integrated luminosity of previous years. The preliminary results of W-boson $A_L$ from 2013 data sample will be presented.
}

\FullConference{XXV International Workshop on Deep-Inelastic Scattering and Related Subjects\\
		3-7 April 2017\\
		University of Birmingham, UK}

\begin{document}

\section{Introduction}

The nucleon spin structure is a fundmental question in QCD and has been studied extensively both in theory and in experiment.  
The quark spin has been found to contribute about 30\% of the nucleon spin and the valence quark helicity distributions are well determined at intermediate $x$ from DIS experiments.
However, the polarization of the sea quarks $\bar u$ and $\bar d$ is not well constrained by SIDIS data \cite{WenChen2014}. 
The production of $W^{+(-)}$ bosons in $p+p$ collisions with longitudinally polarized beams  provides an unique tool to access sea quark and valence quark helicity distributions without the need for hadron fragmentation functions.
As $W$ production is a parity-violating process, $W$ bosons couple only to either left-handed particles or right-handed antiparticles, thus selecting only the corresponding helicity-negative quark and helicity-positive antiquark from the polarized proton \cite{Bunce00}. 

The longitudinal single-spin asymmetry $A_L$ for $W$ boson production with one polarized proton beam is defined as:
 \begin{equation}
A_L(W)\equiv \frac {\sigma{(p_+p \to W)}-\sigma{(p_-p \to W)}}
{\sigma{(p_+p \to W)}+\sigma{(p_-p \to W)}},
 \label{eqa:alw}
\end{equation}
where $\sigma(p_\pm p \to W)$ is the $W$ cross section with a helicity positive/negative proton beam, with subscripts ``$\pm$'' here denoting the helicity.
As the production of $W$ bosons violates parity maximally,  the above $A_L$ equation for $W^\pm$ 
can be rewritten as follows at leading order:
\begin{equation}
A_L^{W^+} = -\frac {\Delta{u(x_1)\bar d(x_2)}-\Delta{\bar
d(x_1)u(x_2)}} {{u(x_1)\bar d(x_2)}+{\bar d(x_1)u(x_2)}},
\label{eqa:alwp}
\end{equation}
\begin{equation}
A_L^{W^-} = -\frac {\Delta{d(x_1)\bar u(x_2)}-\Delta{\bar
u(x_1)d(x_2)}} {{d(x_1)\bar u(x_2)}+{\bar u(x_1)d(x_2)}}.
\label{eqa:alwm}
\end{equation}
$A_L^{W^+}$ approaches -$\Delta u/u$ in the limit $y_W \gg 0$, as the rapidity along the polarized proton beam is defined as positive, and the valence quark usually carries a larger momentum fraction than a sea quark. 
Similarly, $A_L^{W^+}$ approaches $\Delta \bar d/\bar d$ when $y_W \ll 0$. $\Delta \bar u/\bar u$ can be accessed in the backward region of $W^-$ similarly.
Thus, by choosing the rapidity of the $W$ bosons,  the polarizations of antiquarks can be accessed directly.
 
In experiments, $W^\pm$ can be detected via the leptonic channel with the kinematics of isolated decay lepton alone. 
A theoretical framework that describes inclusive lepton production from $W$ boson decay has been developed, so the measurements of $A_L$ versus the lepton rapidity can be compared with theoretical predictions \cite{RHICBOS,CHE}.
The first measurements of the spin asymmetry, $A_L$, and cross section for $W^\pm$ production were reported by the STAR~\cite{Aggarwal:2010vc,STAR:2011aa} and PHENIX~\cite{Adare2011} experiments using data taken in 2009 at $\sqrt{s}=500$~GeV.

In 2012, the STAR experiment collected $pp$ data with an integrated luminosity of about 72~pb$^{-1}$ at $\sqrt{s}=510$~GeV and an average beam polarization of about 56\%, which is much larger than the 2009 data sample, allowing a pseudorapidity-dependent measurement of $A_L$ for $W$ bosons.
In 2011, STAR also collected a relatively small data sample with an integrated luminosity of about 11 pb$^{-1}$ with a similar beam polarization. 
The $A_L$ results from the STAR 2011+2012 data \cite{Adamczyk:2014xyw}  provided new constraints for $\Delta \bar u$ and $\Delta \bar d$ compared to those from previous data, and indicated for the first time that $\Delta \bar u$ is positive, as shown in NNPDF~\cite{NNPDF14} and DSSV++ \cite{Aschenauer:2013woa} global analysis.
These results clearly suggest that the flavor symmetry is also broken in the polarized light sea quark, where $\Delta \bar u$ is positive and $\Delta \bar d$ is negative.
In 2013, STAR collected a much larger data sample, with an integrated luminosity of about 300 pb$^{-1}$ at $\sqrt s$= 510 GeV with an average beam polarization of about 56\%. 
This is more than three times larger than the total integrated luminosity of previous years. 
In this contribution, we report new preliminary results on W single spin asymmetry from STAR data taken in 2013.

\section{Experiment and analysis}
At the STAR experiment, $W^{+(-)}$ bosons are detected through $e^{+(-)}$ with large transverse momenta via the $W \rightarrow e\nu$ channel with the Electromagnetic Calorimeter. 
The $W \rightarrow e\nu$ events are characterized by an isolated $e^{\pm}$ with a sizable transverse energy, $E_T^e$, that peaks near half the $W$ mass ($\sim$40 GeV, referred to as the Jacobian peak).
The neutrino from $W$ decay, close to opposite in the azimuthal of decay $e^{\pm}$, with a large missing transverse energy, is undetected, which causes a large imbalance in the $p_T$ vector sum of all the reconstructed final state particles.
In contrast, the $p_T$ vector sum is well balanced for background events such as $Z/\gamma^* \to e^+e^-$ and QCD di-jet or multi-jet events.
The key selection cuts for $W$ signals are based on the electron isolation and $p_T$ imbalance features.
The $e^+$ and $e^-$ charge separation is done using the Time Projection Chamber (TPC) at mid-rapidity, which provides momenta and charge sign information for charged particles, covering the full azimuth and a pseudorapidity range of  -1.3$ < \eta <$ 1.3.
The Barrel and Endcap Electromagnetic Calorimeters (BEMC and EEMC) cover full azimuth and pseudorapidity ranges of $-1 < \eta < 1$ and 1.1 < $\eta$ < 2.0 respectively.

The $W$ selection starts with the electron isolation cut. First, EMC energy clusters are reconstructed from $2\times2$ calorimeter towers pointed by the extrapolated tracks with high transverse momenta ($p_T$ > 10 GeV). The cluster energy $E_T^{2\times 2}$ in these towers is assigned to the candidate electron. 
Then,  further stages of isolation cuts are implemented, where selection criteria are slightly different between BEMC and EEMC regions.
First it is required that the transverse energy ratio in this 2$\times$ 2 cluster over a surrounding 4$\times$4 cluster is larger than 95\% (96\%) in the BEMC (EEMC) region. 
It is further required that the candidate electron carry a large fraction ( larger than 88\% for both BEMC and EEMC) of the energy in the near-side cone with radius $\Delta R=0.7$, where $\Delta R=\sqrt{\Delta \phi ^2+\Delta \eta ^2}$. 
After that, a $p_T$-balance variable is calculated from the $p_T$ vector sum of the electronc candidate and the reconstructed jets outside an isolation cone around the electron track with a radius $\Delta R = 0.7$, which is then projected to the direction of the candidate electron (named by ``signed $p_T$-balance'').
The signed $p_T$-balance is required to be larger than 14 GeV (20 GeV) for the BEMC (EEMC) region. 
Here, the jet reconstruction use the standard anti-$k_T$ algorithm \cite{anKt}. 
In addition, the total transverse energy in the azimuthally away side cone is required to be less than 11 GeV to further suppress backgrounds in the BEMC region. 
For the EEMC region, another isolation ratio from energy deposit in the EEMC shower maximum detector (ESMD) is also used ($R_{ESMD}>0.7$). 

\begin{figure}[h]
\begin{center}
\includegraphics[width=150mm]{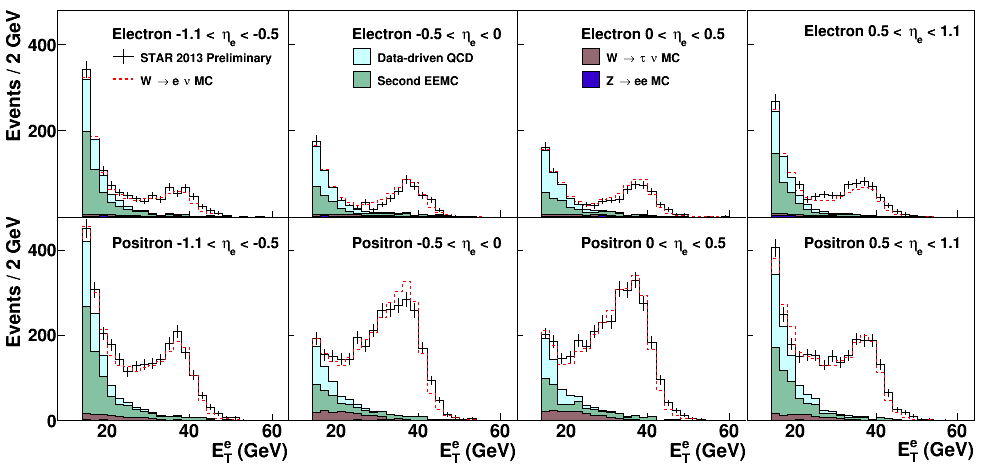}
\caption{
(color online) $E_T^e$ distribution for $W^-$ (top row) and $W^+$ (bottom row) candidates (black), background contributions, and the sum of background and $W\rightarrow e+\nu_e$ Monte Carlo (MC) signals (red dashed) from STAR 2013 $p+p$ collisions\cite{Jinlong,Devika}.
}
\label{fig:Wmass2013}
\end{center}
\end{figure}

\begin{figure}[h]
\begin{center}
\includegraphics[width=140mm]{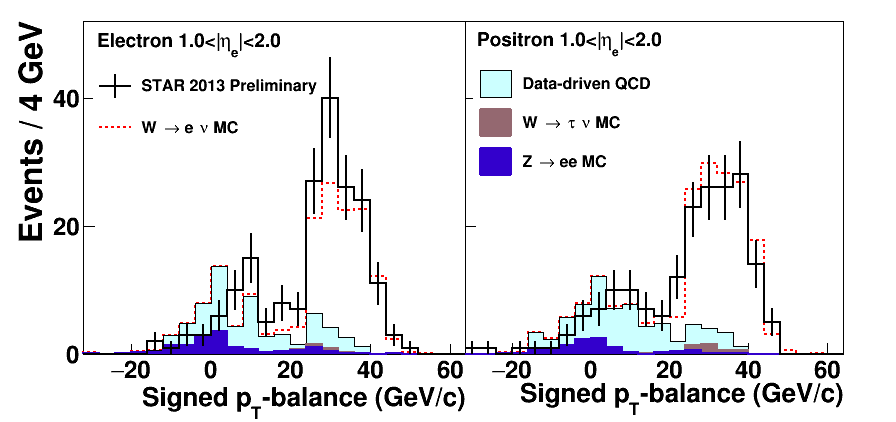}
\caption{
(color online) Signed $p_{T}$-balance distributions for $W^+$ (left) and $W^-$ (right) candidate events (black), background contributions, and $W \rightarrow e\nu$ MC signal (red-dashed) in the EEMC region from STAR 2013 $p+p$ collisions.
}
\label{fig:Wmass2013eemc}
\end{center}
\end{figure}

There are several contributions to the residual background events under the W signal peak (see Fig. \ref{fig:Wmass2013}). 
A di-jet event or $Z/\gamma^* \to e^+e^-$ event, can have one of its jets or electrons outside the STAR acceptance. Such events can be accepted if the detected jet or electron passes all the above W selection criteria,  and the former case is usually referred as "QCD background".
In addition, the W boson can decay to $\tau + \nu$ and $\tau$ can further decay to an electron and a neutrino, though we do not distinguish this feed down contribution.
To estimate the $Z/\gamma^*$ and $\tau$ contributions, Monte Carlo (MC) events are generated using the PYTHIA \cite{PYTHIA} generator and propagated through the STAR detector simulation framework based on GEANT3.
Then the MC events are embedded into STAR zero-bias $p+p$ events and analyzed with the same analysis algorithms.
The QCD background is estimated first using the existing EEMC detector for the uninstrumented acceptance region on the opposite side of the collision point and the rest is estimated using a data-driven method. 
In Fig. \ref{fig:Wmass2013}, the $E_T$ distributions for $W^+$ and $W^-$ are shown for different pseudorapidity intervals covered by BEMC, where the black histograms are the raw signal and the colored histograms are for different background contributions mentioned above.
The residual background fraction is found to be a few percent and is corrected in the determination of single spin asymmetry for $W$ signals, as detailed in Ref.\cite{Adamczyk:2014xyw}.
Figure \ref{fig:Wmass2013eemc} shows the signed $p_T$-balance distribution for $W^+$ and $W^-$ candidates in the near-forward region with EEMC. 

  \section{Results and discussion}
  
The spin-sorted $W$ yields are calculated in the signal window, $25 < E_T < 50$ GeV for each pseudorapidity interval, and the longitudinal single-spin asymmetry $A_L$ is extracted using the following equation:
\begin{equation}
A_L(W)= \frac{1}{\beta}\frac{1}{P}\frac {N_+/l_+ - N_-/l_-}
{N_+/l_+ + N_-/l_-},
 \label{ALbeta}
\end{equation}
where $\beta$ quantifies the dilution due to residual background, $P$ is the beam polarization, $N_+(N_−)$ is the W yield when the helicity of the polarized beam is positive (negative), and $l_\pm$ are the relative luminosity correction factors. 
The relative luminosity factors are calcuated independently with non-W events with high precision.

  \begin{figure}[t]
 \begin{center}
 \includegraphics[width=0.65\textwidth]{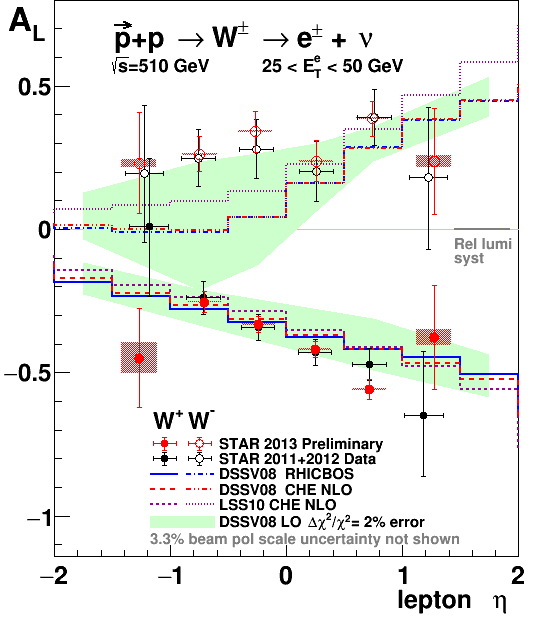} %0.55
 \caption{ Single-spin asymmetry $A_L$ for $W^\pm$ production as a function of $\eta_e$  in $p+p$ collisions at 510 GeV at STAR using 2013 data, in comparison to published 2011+2012 data\cite{Adamczyk:2014xyw}.}
 \label{Fig:figure2}
 \vspace*{-5mm}
 \end{center}
\end{figure}
 
 \begin{figure}[t]
 \begin{center}
 %\includegraphics[width=0.8\textwidth]{moneyPlot2013_wtPRL12.png} %0.55
 %\caption{ Single-spin asymmetry $A_L$ for $W^\pm$ production as a function of $\eta_e$  in $pp$ collisions at 510 GeV at PHENIX using 2011, 2012, and 2013 data \cite{Adare:2015gsd}.}
 \label{Fig:pi0ALL}
 \vspace*{-5mm}
 \end{center}
\end{figure}

Figure 3 shows the preliminary results of  $A_L$ for $W^\pm$ boson at six lepton pseudorapidity bins from STAR 2013 data sample \cite{Jinlong,Devika}  in comparison with theoretical predictions and previous STAR results from the  2011+2012 data \cite{Adamczyk:2014xyw}. 
The 2013 preliminary $A_L$ results are consistent with the published 2011+2012 results, with 40\% reduced statistical uncertainties. 
The $A_L$ data for $W^+$ remain to be consistent with the theoretical predictions, for example the predictions using DSSV08 \cite{DSSV08} and LSS10 \cite{LSS10} global analysis.  
However, the measured $A_L$ for $W^-$ are systematically larger than the theoretical predictions, in the region of $\eta_{e^-}$ $<$ 0, which continue to support a positive $\Delta \bar u$ as 2011+2012 results. 
%$A_L$ for $W^+$ remains to be consistent with the theoretical predictions albeit decreasing slightly more steeply with increasing pseudorapidity. 
%AL for W−, however, are above the theoretical predictions at negative pseudorapidity. 
The 2011+2012 results have been included into the global QCD analysis by the NNPDF group\cite{NNPDF14}, which suggests a symmetry breaking for $\Delta \bar u$ and $\Delta \bar d$.
The constraints provided by these STAR data lead to a shift in the central value of $\Delta \bar u$ from negative to positive in x range 0.05 < x < 0.2. 
The recent DSSV$^{++}$ global analysis has also included the STAR 2012 data, and reported a similar impact on the $\Delta \bar u$ and $\Delta \bar d$ distributions \cite{Aschenauer:2013woa}. 
The STAR 2013 new results show its impact in a re-weighted NNPDF fit \cite{NNPDF16}, which provide further constraints on light sea quark polarization.
STAR 2013 W $A_L$ results have reached unprecedented precision and are expected to significantly advance our understanding of nucleon spin structure.

The author is supported partially by the MoST of China (973 program No. 2014CB845400), the Natural Science Foundation of Shandong Province, China, under Grant No.ZR2013JQ001.

\end{document}